\documentclass[12pt]{article}
 
\usepackage[utf8]{inputenc}
\usepackage{graphicx}
\usepackage{float}
\usepackage{amsmath,amssymb,amsthm}
\usepackage{mathrsfs}
\usepackage{bm}
\usepackage{braket}
\usepackage{framed}
\usepackage{color}
\usepackage{comment}
\usepackage{ulem}
\usepackage{tcolorbox}
\usepackage[hang,small,bf]{caption}
\usepackage[subrefformat=parens]{subcaption}
\graphicspath{{./}}

\makeatletter

\renewcommand\section{\@startsection {section}{1}{\z@}%
                                   {-3.5ex \@plus -1ex \@minus -.2ex}%
                                   {2.3ex \@plus.2ex}%
                                   {\normalfont\large\bfseries}}

\renewcommand\subsection{\@startsection{subsection}{2}{\z@}%
                                     {-3.25ex\@plus -1ex \@minus -.2ex}%
                                     {1.5ex \@plus .2ex}%
                                     {\normalfont\normalsize\bfseries}}

\makeatother

\begin{document}

\baselineskip=18pt
\numberwithin{equation}{section}
\allowdisplaybreaks


\thispagestyle{empty}

\vspace*{-2cm}
\begin{flushright}
\end{flushright}

\begin{center}
\vspace{1.4cm}

{\bf \Large Higher-Form Description of Chiral Soliton Lattice}

\vspace{1.3cm}

{\bf Yutaka Ookouchi} \\
\vspace*{0.5cm}

{\it Faculty of Arts and Science, Kyushu University, Fukuoka 819-0395, Japan}\\

\vspace*{0.5cm}
\end{center}

\vspace{1cm} \centerline{\bf Abstract} \vspace*{0.5cm}

The chiral soliton lattice (CSL) is conventionally described as a periodic winding configuration of a compact scalar field.  In this paper, we formulate the CSL in a higher-form framework based on a St\"uckelberg coupling between a two-form gauge field $C_2$ and a three-form gauge field $A_3$, in which strings and walls can be treated directly as extended sources.  Higher-form gauge invariance requires the string worldsheet to be the boundary of the wall worldvolume.  In the quadratic dual effective theory, exchange of the massive higher-form mode generates a repulsive Yukawa interaction between parallel walls.  With its coefficient fixed by duality, the resulting CSL lattice spacing agrees well with the exact Sine--Gordon result without parameter fitting.  We then extend the description to a nonlinear, multi-branch higher-form theory.  It reproduces the exact dilute wall--wall interaction coefficient, the periodic CSL and its energy-minimization condition, as well as the gapless CSL phonon.  The phonon is identified as a collective displacement of the periodic higher-form background, while scalar winding is reinterpreted as branch flow of the nonlinear higher-form theory.

\newpage
\setcounter{page}{1}


\section{Introduction}

Topological solitons provide a common language for nontrivial phases in quantum field theory and condensed-matter systems.  A particularly instructive example is the chiral soliton lattice (CSL), a spatially periodic state built from repeated winding of a compact field and familiar from chiral magnets and related systems \cite{Dzyaloshinskii:1964,Togawa:2016,Kishine:2015}.  In QCD, anomalous interactions in dense matter endow neutral-pion domain walls with baryon quantum numbers and can make a stack of such walls energetically favorable in a sufficiently strong magnetic field \cite{Son:2004tq,Son:2007ny}.  The exact periodic solution was subsequently constructed, establishing the neutral-pion CSL as a ground state of QCD at nonzero baryon chemical potential and strong magnetic field within a controlled low-energy effective theory \cite{Brauner:2016pko}.  Each unit cell carries baryon charge and magnetic moment.  Related CSL phases have also been studied at nonzero temperature, under rotation, and in QCD-like theories \cite{Brauner:2017uiu,Brauner:2021sbn,Huang:2017pqe,Nishimura:2020odq,Brauner:2019rjg,Brauner:2019fnl,Yamada:2021ntd}.

The conventional low-energy description of the CSL is remarkably economical.  A compact pseudo-Nambu--Goldstone field has a periodic potential, while the anomaly supplies a total derivative that lowers the energy of configurations with nonzero winding.  The local wall profile is governed by the Sine--Gordon equation, whereas the external chemical potential and magnetic field determine whether adding a unit of winding is energetically favorable.  Near the transition to the CSL phase, the lattice spacing is large and each soliton is well separated from its neighbors.  As the magnetic field is increased, the solitons move closer together and eventually form a smooth periodic profile \cite{Brauner:2016pko,Higaki:2022qwt}.  The scalar formulation therefore gives an exact description of the ground state.  However, it is less convenient for questions in which the extended objects themselves are the natural dynamical variables: interactions among separated walls, finite walls ending on strings, and configurations whose wall and string geometries evolve together.  This motivates a complementary formulation in which walls and strings enter directly as extended sources.

Such a description is provided by higher-form gauge fields.  The effective description of axion-like defects as strings and open or closed membranes has a long history \cite{Townsend:1993}.  More recently, finite wall disks bounded by string loops have been used to study CSL formation in both Nambu--Goto-type and microscopic descriptions \cite{Higaki:2022qwt,Eto:2022lcs}.  In parallel, generalized-symmetry and higher-form approaches have clarified the role of compact-scalar winding, domain walls, and their associated gauge fields \cite{Banks:2010zn,Gaiotto:2014kfa,Hidaka:2019threeform,Hidaka:2020axion3group,Brennan:2020e}.  A massive axion can be dualized to a St\"uckelberg system of two- and three-form gauge fields, and nonlinear functions of the four-form field strength can encode nonquadratic axion potentials, including a cosine potential \cite{Dvali:2005}.  The global structure of such systems, including gauge-invariant wall--string junctions, has also been analyzed systematically \cite{Anber:2024}.

The higher-form formulation has both a conceptual and a practical advantage for the present problem.  Conceptually, the goal is to describe the CSL directly in terms of its constituent extended objects, while preserving the quantitative information provided by the exact scalar solution.  Practically, a St\"uckelberg system of two- and three-form gauge fields provides a common gauge-theoretic description of a wall and its boundary string.  Higher-form gauge invariance relates the two objects and requires the string worldsheet to be the boundary of the wall worldvolume.  This framework is particularly well suited to finite wall--string composites, in which both the wall and the string can change their shapes dynamically.  A simple example already appears in the present work, where higher-form exchange generates not only interactions between separated walls but also a finite-size self-interaction for a string loop bounding a wall disk.  More generally, the same framework provides a natural starting point for dynamical processes involving the creation or evolution of wall--string configurations.  A detailed study of such nucleation processes within the higher-form framework will be presented in a separate work.

The first question we address is how much of the CSL physics is already captured by the minimal quadratic higher-form theory.  We consider one two-form field $C_2$ and one three-form field $A_3$, with the dual couplings fixed by the Sine--Gordon normalization.  Exchange of the resulting massive higher-form mode produces a repulsive Yukawa interaction between equally oriented walls.  Summing this interaction over a periodic array gives an estimate of the CSL lattice spacing that closely follows the exact Sine--Gordon result over a broad range of magnetic field.  The same quadratic theory can also be used to study finite wall--string configurations, including the self-interaction of a string loop bounding a wall disk.

The quadratic theory is nevertheless not the full dual of the Sine--Gordon model.  In the dilute regime it reproduces the exponential range of the wall--wall interaction but misses its exact coefficient by an order-one factor.  This discrepancy provides a useful diagnostic: the long-distance mass scale is already present in the quadratic theory, whereas the normalization retains nonlinear information about the soliton core and its comparison to the tail.  We therefore extend the quadratic theory to a nonlinear, multi-branch higher-form theory.  The resulting theory reconstructs the exact periodic CSL and its energy-minimization condition.  Its stress tensor reproduces the exact dilute wall--wall interaction coefficient, and linear fluctuations around the periodic higher-form background reproduce the gapless CSL phonon and its anisotropic low-energy dispersion.

The nonlinear completion also reveals a conceptual aspect that is hidden in a purely local dualization.  In the scalar formulation, the CSL repeatedly winds through the compact field space.  In the higher-form formulation, this repeated winding is encoded as flow between neighboring branches of the nonlinear theory.  Likewise, the CSL phonon becomes a collective displacement of the periodic higher-form background.  Thus the higher-form description is not merely a change of variables for a single wall: its global branch structure keeps track of repeated winding, while its gauge-field variables organize the interactions and dynamics of the extended defects.  The CSL therefore provides a concrete example in which higher-form duality can be used not only to describe the topology of walls and strings, but also to calculate their interactions and collective dynamics.

The nonlinear three-form description of a periodic scalar potential, as well as the higher-form structure of axion wall--string systems, has been developed in earlier work \cite{Dvali:2005,Anber:2024}.  In particular, Ref.~\cite{Anber:2024} formulated the nonlinear three-form theory and its Kalb--Ramond dual, including domain walls and their attachment to axion strings.  Here we apply this framework to the chiral soliton lattice, for which the exact Sine--Gordon solution provides a quantitative reference, with particular attention to the branch flow associated with repeated winding.

The paper is organized as follows.  Section~2 reviews the Sine--Gordon CSL and derives the minimal quadratic higher-form formulation, including the gauge-invariant wall--string source and the normalization of the dual fields.  Section~3 derives the defect interaction generated by the massive higher-form mode and illustrates the finite wall--string disk, carefully separating local tension renormalization from finite-size effects.  Section~4 applies the quadratic interaction to the CSL and compares the resulting lattice spacing with the exact solution.  Section~5 develops the nonlinear multi-branch completion and reconstructs the exact CSL as a branch-flow configuration.  Section~6 studies the dynamics of this nonlinear higher-form CSL, deriving the exact dilute wall interaction and the gapless phonon as a collective displacement of the periodic higher-form background.  Section~7 summarizes the results and discusses extensions.

\section{Minimal higher-form formulation}

\subsection{Sine--Gordon description of the CSL}

We begin with the low-energy model used in analyses of CSL formation and related wall--string dynamics \cite{Brauner:2016pko,Higaki:2022qwt,Eto:2022lcs},
\begin{equation}
 {\cal L}_{\rm SG}
 =\frac{f^2}{2}(\partial_\mu\phi)^2
 +f^2m^2(\cos\phi-1)
 +\frac{\mu}{4\pi^2}{\bm B}\cdot\nabla\phi,
 \qquad \phi\simeq\phi+2\pi .
 \label{eq:SGaction}
\end{equation}
Here $f$ is the decay constant of the compact field, $m$ is its Sine--Gordon mass, $\mu$ is the baryon chemical potential, and ${\bm B}$ is an external magnetic field.  For ${\bm B}=B\hat{\bm z}$ and a static configuration depending only on $z$, the last term is a total derivative and the local equation of motion is
\begin{equation}
 \phi''(z)=m^2\sin\phi(z).
 \label{eq:SGeom}
\end{equation}
The isolated wall interpolating between neighboring vacua related by the \(2\pi\) periodicity of the compact scalar is
\begin{equation}
 \phi_{\rm DW}(z)=4\tan^{-1}e^{mz},
 \qquad
 \phi(-\infty)=0,
 \qquad
 \phi(+\infty)=2\pi .
 \label{eq:SGwall}
\end{equation}
Its positive intrinsic Sine--Gordon tension, before the topological derivative term is included, is
\begin{equation}
 \sigma_0=8mf^2.
 \label{eq:sigma0}
\end{equation}
For one $2\pi$ winding the total derivative lowers the static energy per unit area by $\mu B/(2\pi)$.  We therefore define the net wall energy per area in the external background by
\begin{equation}
 \sigma_{\rm phys}
 \equiv \sigma_0-\frac{\mu B}{2\pi}
 =8mf^2(1-\beta),
 \qquad
 \beta\equiv\frac{\mu B}{16\pi mf^2}.
 \label{eq:sigmaphys}
\end{equation}
Here $\sigma_0$ denotes the positive local wall tension, while $\sigma_{\rm phys}$ includes the topological energy gain.  Thus $\sigma_{\rm phys}=0$ at the point where the CSL first appears, $\beta=1$ and becomes negative for $\beta>1$.  A negative $\sigma_{\rm phys}$ does not mean that the microscopic wall has a negative local tension; it means that, in the fixed external $(\mu,B)$ background, the formation of domain walls lowers the total energy.

The periodic solution of \eqref{eq:SGeom} can be written as
\begin{equation}
 \phi(z)=\pi+2\,\mathrm{am}\left(\frac{mz}{k},k\right),
 \label{eq:exactCSL}
\end{equation}
where $0<k<1$ is the elliptic modulus, whose value is determined by minimizing the energy in the external $(\mu,B)$ background, and $\mathrm{am}$ is the Jacobi amplitude.  Its period is
\begin{equation}
 \ell(k)=\frac{2kK(k)}{m},
 \label{eq:exactperiod}
\end{equation}
where $K(k)$ is the complete elliptic integral of the first kind.  The ground-state value of $k$ is determined by \cite{Brauner:2016pko,Higaki:2022qwt}
\begin{equation}
 \frac{E(k)}{k}=\beta,
 \label{eq:exactcondition}
\end{equation}
with $E(k)$ the complete elliptic integral of the second kind.  Since $E(k)/k\geq1$, the CSL first appears at $\beta=1$.  As $\beta\to1$, $k\to1$ and $m\ell\to\infty$, so the CSL is a dilute lattice of well-separated walls.  At larger $\beta$ the spacing decreases and the profile becomes increasingly smooth.  We use $\operatorname{sn}$, $\operatorname{cn}$, and $\operatorname{dn}$ below for the standard Jacobi elliptic functions with the same modulus $k$.

\subsection{Wall--string sources and higher-form gauge invariance}

We now reformulate the wall--string system in terms of higher-form gauge fields, so that the wall and its boundary string can be treated directly as extended objects. We introduce a two-form gauge field $C_2$ and a three-form gauge field $A_3$, and define the gauge-invariant three-form
\begin{equation}
 H_3=dC_2-A_3,
 \label{eq:H3def}
\end{equation}
with higher-form gauge transformations
\begin{equation}
 C_2\longrightarrow C_2+\Lambda_2,
 \qquad
 A_3\longrightarrow A_3+d\Lambda_2.
 \label{eq:HFgauge}
\end{equation}
In theories with a massive periodic scalar, a finite domain wall can end on a string, and such wall--string configurations can be described at low energies by a Nambu--Goto-type action \cite{Higaki:2022qwt,Townsend:1993,Eto:2022lcs}.  The role of the higher-form formulation is to encode the same attachment condition directly in the gauge structure of the source.  Let $W_3$ denote the wall worldvolume and $\Sigma_2$ the string worldsheet.  At leading derivative order their geometrical actions are
\begin{equation}
 S_{\rm NG}
 =-\sigma_0\int_{W_3}d^3\zeta\sqrt{\det h}
  -T\int_{\Sigma_2}d^2\xi\sqrt{-\det\gamma}.
 \label{eq:NG}
\end{equation}
Here $h_{ab}$ and $\gamma_{ij}$ are the induced metrics on $W_3$ and $\Sigma_2$, while $\zeta^a$ and $\xi^i$ are their worldvolume coordinates.  The local coefficients $\sigma_0>0$ and $T>0$ are the intrinsic wall and string tensions.  When the external $(\mu,B)$ background is included in a static energy, the wall area term is instead governed by the net quantity $\sigma_{\rm phys}$ defined in \eqref{eq:sigmaphys}.

A three-form gauge potential couples electrically to a wall, while a two-form couples electrically to a string.  In the minimal normalization relevant here we take
\begin{equation}
 S_{\rm src}
 =\int_{\Sigma_2}C_2-\int_{W_3}A_3.
 \label{eq:sourceexplicit}
\end{equation}
The higher-form gauge transformation \eqref{eq:HFgauge} gives
\begin{equation}
 \delta S_{\rm src}
 =\int_{\Sigma_2}\Lambda_2-\int_{W_3}d\Lambda_2.
\end{equation}
Hence gauge invariance requires
\begin{equation}
 \partial W_3=\Sigma_2,
 \label{eq:boundarycondition}
\end{equation}
and by Stokes' theorem the two source terms combine into
\begin{equation}
 S_{\rm src}=\int_{W_3}(dC_2-A_3)=\int_{W_3}H_3.
 \label{eq:sourceH}
\end{equation}
The attachment of the string to the wall is therefore not an additional geometric rule imposed after writing the action.  It is the condition required for the extended source to be compatible with the gauge symmetry.  Gauge-invariant wall--string operators of this general type also arise in modern analyses of axion--three-form systems \cite{Anber:2024}.

\subsection{Scalar--two-form duality and normalization}

Following the three-form formulation of axion dynamics \cite{Dvali:2005,Anber:2024}, we consider
\begin{equation}
 S[\phi,A_3]
 =\frac{f^2}{2}\int d\phi\wedge *d\phi
 -\frac{1}{2\pi}\int \phi\,dA_3 .
 \label{eq:scalarA3}
\end{equation}
After integrating the second term by parts and replacing $d\phi$ by an independent one-form $V_1$, a convenient first-order action is
\begin{equation}
 S_{\rm 1st}
 =\frac{f^2}{2}\int V_1\wedge *V_1
 -\frac{1}{2\pi}\int C_2\wedge dV_1
 +\frac{1}{2\pi}\int V_1\wedge A_3 .
 \label{eq:firstorder}
\end{equation}
The $C_2$ equation imposes $dV_1=0$, locally giving $V_1=d\phi$.  Integrating the $C_2\wedge dV_1$ term by parts, the $V_1$ equation instead gives
\begin{equation}
 f^2 *V_1-\frac{1}{2\pi}H_3=0.
 \label{eq:V1relation}
\end{equation}
Together with $V_1=d\phi$, this gives the duality relation
\begin{equation}
 H_3=2\pi f^2*d\phi .
 \label{eq:dualityrelation}
\end{equation}
Eliminating $V_1$ yields
\begin{equation}
 S_{C_2}
 =\frac{1}{8\pi^2f^2}\int H_3\wedge *H_3
 \equiv
 \frac{1}{2g^2}\int H_3\wedge *H_3,
 \label{eq:C2kin}
\end{equation}
where $g=2\pi f$.
The coefficient of the dual kinetic term is thus fixed by the normalization of the compact scalar.

\subsection{Quadratic St\"uckelberg completion}

A massive three-form obtained through this type of St\"uckelberg mechanism is familiar from three-form formulations of axion physics \cite{Dvali:2005}.  Here our purpose is to fix the parameters of that quadratic theory from the Sine--Gordon description and then use the resulting massive mode as the mediator of the long-distance interaction between CSL walls.  We now include a kinetic term for the three-form gauge field $A_3$.
The quadratic bulk theory with the wall--string source takes the form
\begin{equation}
 S_{\rm quad}
 =S_{\rm NG}
 +\frac{1}{2g^2}\int H_3\wedge *H_3
 -\frac{1}{2e^2}\int F_4\wedge *F_4
 +\int_{W_3}H_3,
 \label{eq:quadfull}
\end{equation}
where $H_3=dC_2-A_3$ and $F_4=dA_3$.
In the local unitary gauge $C_2=0$, this is a massive three-form theory.  The mass is
\begin{equation}
 m_A=\frac{e}{g}.
 \label{eq:mA}
\end{equation}
Identifying this pole with the Sine--Gordon mass gives
\begin{equation}
 g=2\pi f,
 \qquad
 m_A=m,
 \qquad
 e=2\pi fm .
 \label{eq:comparison}
\end{equation}
Together with the unit source normalization, \eqref{eq:comparison} leaves no continuous coefficient that can be adjusted to fit the wall--wall interaction.  The quadratic theory should therefore be viewed as a predictive truncation of the dual description.

\section{Defect interactions in the quadratic dual theory}

\subsection{Induced current action}

The interaction mediated by the St\"uckelberg sector is most transparent after dualizing the three-form field strength $H_3$ to a vector,
\begin{equation}
 h^\mu\equiv\frac{1}{3!}\epsilon^{\mu\nu\rho\sigma}H_{\nu\rho\sigma}.
 \label{eq:hvector}
\end{equation}
where $\epsilon^{\mu\nu\rho\sigma}$ is the Levi--Civita tensor. We choose the orientation $\epsilon_{0123}=+1$ in flat spacetime. With the metric signature $(+,-,-,-)$, this implies $\epsilon^{0123}=-1$. For a wall current $j^\mu$, the quadratic action may be written schematically as
\begin{equation}
 S[h]
 =\int d^4x\left[
 -\frac{1}{2g^2}h_\mu h^\mu
 +\frac{1}{2e^2}(\partial_\mu h^\mu)^2
 -h_\mu j^\mu
 \right].
 \label{eq:haction}
\end{equation}
Notice that $F_4=-dH_3$, so the $F_4$ kinetic term becomes the longitudinal term $(\partial_\mu h^\mu)^2$. Thus $h^\mu$ does not describe independent transverse propagating modes; the St\"uckelberg system contains a single massive mode, corresponding to the original scalar degree of freedom.
Integrating out $h_\mu$ gives the current--current action
\begin{equation}
 S_{\rm ind}[j]
 = \frac{g^2}{2}\int\frac{d^4k}{(2\pi)^4}
 j_\mu(-k)
 \left[
 \eta^{\mu\nu}
 -\frac{k^\mu k^\nu}{k^2-m_A^2+i\epsilon}
 \right]j_\nu(k).
 \label{eq:inducedaction}
\end{equation}
The pole at $k^2=m_A^2$ corresponds to the massive higher-form mode that mediates the interaction between defects. For a configuration containing several walls, the total current is the sum of the individual wall currents, $j^\mu=\sum_i j_i^\mu$, where $j_i^\mu$ denotes the current associated with the $i$-th wall. Substituting this decomposition into \eqref{eq:inducedaction} separates the induced action into self-energy terms and pairwise interaction terms. The self-energy terms contain local renormalizations as well as finite contributions that depend on the size and shape of the defect, whereas the pairwise terms determine the separation-dependent force between distinct walls.

\subsection{Parallel walls and the Yukawa force}

Place two infinite parallel walls at $z=0$ and $z=\ell$, and denote
their transverse area by $A$.  Let $s_i=\pm1$ denote the orientation
of the $i$-th wall.  Their currents have only a component normal to
the walls,
\[
 j_i^z(x)=s_i\,\delta(z-z_i),
 \qquad z_1=0,\qquad z_2=\ell .
\]
For an infinite wall, there is no boundary and hence no associated string source.

Substituting $j^\mu=j_1^\mu+j_2^\mu$ into
\eqref{eq:inducedaction}, the cross term gives the interaction
between the two walls.  For this static and translationally invariant
configuration, $k_0=k_x=k_y=0$, and the $zz$ component of the kernel
reduces to
\[
 \eta^{zz}
 -\frac{k_z^2}{k^2-m_A^2+i\epsilon}
 =
 -1+\frac{k_z^2}{k_z^2+m_A^2}
 =
 -\frac{m_A^2}{k_z^2+m_A^2}.
\]
Using $S_{12}=-\Delta t\,V_{12}$, where $\Delta t$ denotes the time interval, the separation-dependent interaction is
therefore
\begin{equation}
 \frac{V_{12}(\ell)}{A}
 =
 g^2s_1s_2
 \int_{-\infty}^{\infty}\frac{dk_z}{2\pi}
 \frac{m_A^2}{k_z^2+m_A^2}e^{ik_z\ell}.
 \label{eq:wallintintegral}
\end{equation}
Using
\begin{equation}
 \int_{-\infty}^{\infty}\frac{dk}{2\pi}
 \frac{e^{ik\ell}}{k^2+m^2}
 =\frac{e^{-m|\ell|}}{2m},
\end{equation}
we obtain
\begin{equation}
 \frac{V_{12}(\ell)}{A}
 =\frac{g^2m_A}{2}s_1s_2e^{-m_A|\ell|}.
 \label{eq:YukawaGeneral}
\end{equation}
Thus equally oriented walls repel, while oppositely oriented walls attract.  After the comparison in \eqref{eq:comparison}, the coefficient for the CSL walls is
\begin{equation}
 C_{\rm dual}=\frac{g^2m}{2}=2\pi^2mf^2.
 \label{eq:Cdual}
\end{equation}
The range of the interaction, $m^{-1}$, coincides with the scale governing the exponential tail of the Sine--Gordon wall.  This is not an additional interaction to be added on top of the exact Sine--Gordon energy.  Rather, it is a quadratic dual representation of the same long-distance wall-tail physics.

\subsection{Finite wall--string disk}

The previous subsection used the cross term in the induced action to determine the interaction between two distinct walls.  We now turn to the self-energy term for a single wall of finite size.  In this case the wall has a boundary, which is accompanied by a string loop in the higher-form source description.  This provides a simple setting in which to examine how the massive higher-form field modifies the energy of a finite wall--string configuration.

Consider a circular wall disk of radius $R$ bounded by a string loop.  We denote by $T_{\rm phys}>0$ the physical string tension, distinguished from the bare string tension $T$ introduced in the Nambu--Goto action.  The leading static Nambu--Goto energy is
\begin{equation}
 E_{\rm NG}^{\rm phys}(R)
 =2\pi T_{\rm phys}R+\pi\sigma_{\rm phys}R^2.
 \label{eq:diskNG}
\end{equation}
In the CSL-favored regime, where $\sigma_{\rm phys}<0$ as defined in \eqref{eq:sigmaphys}, the Nambu--Goto energy \eqref{eq:diskNG} has a barrier.  It is useful to define the position of its maximum by
\begin{equation}
 R_0\equiv-\frac{T_{\rm phys}}{\sigma_{\rm phys}}
 =\frac{T_{\rm phys}}{|\sigma_{\rm phys}|},
 \label{eq:R0disk}
\end{equation}
By construction, $R_0$ is the position of the maximum of the Nambu--Goto barrier, while its nontrivial zero lies at $R=2R_0$.

For a thin disk in the $z=0$ plane, define $r=\sqrt{x^2+y^2}$ and $k_\perp=\sqrt{k_x^2+k_y^2}$.  The Poincare-dual wall current in position space can be written as
\begin{equation}
 j_z({\bm x})=\delta(z)\Theta(R-r).
 \label{eq:diskcurrentx}
\end{equation}
The Heaviside step function $\Theta(R-r)$ ensures that the current is nonzero only inside the disk $r<R$.

Using rotational symmetry in the transverse plane, the Fourier transform can be evaluated in terms of a Bessel function.  Choosing the transverse momentum along the $x$-axis, so that ${\bm k}_\perp\cdot{\bm x}_\perp=k_\perp r\cos\theta$, we find
\begin{align}
 j_z({\bm k})
 &=\int d^2x_\perp\,
 e^{-i{\bm k}_\perp\cdot{\bm x}_\perp}j_z({\bm x})
 =\int_{r<R}d^2x_\perp\,
 e^{-i{\bm k}_\perp\cdot{\bm x}_\perp}
 \nonumber\\
 &=\int_0^R r\,dr\int_0^{2\pi}d\theta\,
 e^{-ik_\perp r\cos\theta}
 =2\pi\int_0^R r\,dr\,J_0(k_\perp r)
\nonumber \\ 
 &=\frac{2\pi R}{k_\perp}J_1(k_\perp R).
 \label{eq:diskcurrent}
\end{align}
where $J_n$ denotes the Bessel function of the first kind.  In obtaining this result, we used the standard integral representation
\begin{equation}
 \int_0^{2\pi}d\theta\,e^{-ia\cos\theta}
 =2\pi J_0(a)
\end{equation}
and the identity
\begin{equation}
 \frac{d}{dx}\left[xJ_1(x)\right]=xJ_0(x).
\end{equation}
Thus the appearance of the Bessel functions is simply a consequence of taking the two-dimensional Fourier transform of a rotationally symmetric disk.  Because the disk has a boundary, the exterior derivative of the Poincare-dual wall one-form current is localized on the circular edge.  This is precisely the current relation dual to $\partial W_3=\Sigma_2$: wall and string sources are not independent ingredients of the theory. 

Evaluating the quadratic response on the disk gives a nonlocal self-energy.  For the static thin source in \eqref{eq:diskcurrent}, the induced energy from the quadratic $C_2$--$A_3$ sector is
\begin{equation}
 E_{\rm ind}(R;m_A)
 =\frac{g^2}{2}
 \int\frac{d^3k}{(2\pi)^3}
 |j_z({\bm k};R)|^2
 \frac{k_\perp^2+m_A^2}
 {k_\perp^2+k_z^2+m_A^2}.
 \label{eq:diskEind}
\end{equation}
For the zero-thickness source, this expression is ultraviolet sensitive because the microscopic structure of the string core is not resolved.  We therefore introduce a transverse-momentum cutoff $\Lambda$, which may be viewed as the inverse core scale.  Performing the $k_z$ integral gives
\begin{equation}
 \int_{-\infty}^{\infty}\frac{dk_z}{2\pi}
 \frac{k_\perp^2+m_A^2}
 {k_z^2+k_\perp^2+m_A^2}
 =\frac{1}{2}\sqrt{k_\perp^2+m_A^2},
 \label{eq:kzintegralcutoff}
\end{equation}
and hence
\begin{equation}
 E_{\rm ind}^{\Lambda}(R;m_A)
 =\frac{\pi g^2R^2}{2}
 \int_0^{\Lambda}dk_\perp\,
 \frac{J_1(k_\perp R)^2}{k_\perp}
 \sqrt{k_\perp^2+m_A^2}.
 \label{eq:diskEcutoff}
\end{equation}
Introducing
\begin{equation}
 u=k_\perp R,\qquad
 \mu_R=m_AR,\qquad
 X=\Lambda R,
\end{equation}
we can write
\begin{equation}
 E_{\rm ind}^{\Lambda}(R;m_A)
 =\frac{\pi g^2R}{2}
 \int_0^Xdu\,
 \frac{J_1(u)^2}{u}\sqrt{u^2+\mu_R^2}.
 \label{eq:diskEdimensionless}
\end{equation}
The large-$R$ structure of this expression separates naturally into contributions associated with the string and the wall.  For this purpose, write
\begin{equation}
 \sqrt{u^2+\mu_R^2}
 =u+\left(\sqrt{u^2+\mu_R^2}-u\right)
\end{equation}
Keeping the finite cutoff $X=\Lambda R$, this decomposition gives the exact expression
\begin{equation}
 E_{\rm ind}^{\Lambda}(R;m_A)
 =\frac{\pi g^2R}{2}
 \left[A(X)+{\cal I}_X(\mu_R)\right],
 \label{eq:diskEsplit}
\end{equation}
where
\begin{align}
 A(X)&\equiv\int_0^Xdu\,J_1(u)^2,
 \label{eq:Adef}\\
 {\cal I}_X(\mu_R)&\equiv\int_0^Xdu\,
 \frac{J_1(u)^2}{u}
 \left(\sqrt{u^2+\mu_R^2}-u\right).
 \label{eq:IXdef}
\end{align}
Thus both integrals have the same upper limit $X$ at this stage.  The second integral, however, has a well-defined $X\to\infty$ limit.  For $u\gg\mu_R$,
\begin{equation}
 \sqrt{u^2+\mu_R^2}-u
 =\frac{\mu_R^2}{2u}
 +O\!\left(\frac{\mu_R^4}{u^3}\right),
\end{equation}
and $J_1(u)^2=O(u^{-1})$.  The integrand in ${\cal I}_X$ therefore falls as $O(\mu_R^2/u^3)$, so that the omitted tail satisfies $O(m_A^2/\Lambda^2)$. For $\Lambda\gg m_A$, this is cutoff suppressed.  We may therefore define the finite function
\begin{equation}
 {\cal I}(\mu_R)\equiv\int_0^\infty du\,
 \frac{J_1(u)^2}{u}
 \left(\sqrt{u^2+\mu_R^2}-u\right),
 \label{eq:Idef}
\end{equation}
and, up to terms suppressed by $m_A^2/\Lambda^2$, write
\begin{equation}
 E_{\rm ind}^{\Lambda}(R;m_A)
 =\frac{\pi g^2R}{2}
 \left[A(\Lambda R)+{\cal I}(m_AR)\right].
 \label{eq:diskEsplitapprox}
\end{equation}
The first function retains the cutoff dependence and has the asymptotic form
\begin{equation}
 A(X)=\frac{1}{\pi}\log X+C_A+o(1),
 \qquad X\to\infty,
 \label{eq:Aasymptotic}
\end{equation}
while the second behaves as
\begin{equation}
 {\cal I}(\mu_R)
 =\frac{\mu_R}{2}
 -\frac{1}{\pi}\log\mu_R
 +C_I+o(1),
 \qquad \mu_R\to\infty.
 \label{eq:Iasymptotic}
\end{equation}
Here $C_A$ and $C_I$ are constants independent of $R$; numerically, $C_A\simeq0.209$ and $C_I\simeq-0.307$.

Combining the large-$R$ expansions in \eqref{eq:Aasymptotic} and \eqref{eq:Iasymptotic}, the logarithmic dependence on $R$ cancels, leaving a term proportional to $R$ that can be absorbed into the physical string tension.  The resulting large-$R$ expansion is
\begin{equation}
 E_{\rm ind}^{\Lambda}(R;m_A)
 =\frac{\pi g^2m_A}{4}R^2
 +\frac{g^2R}{2}\log\frac{\Lambda}{m_A}
 +\frac{\pi g^2}{2}(C_A+C_I)R
 +o(R).
 \label{eq:diskElargeR}
\end{equation}
The area and circumference terms give local contributions to the planar wall tension and the string tension, respectively,
\begin{equation}
 \Delta\sigma=\frac{g^2m_A}{4},
 \qquad
 \Delta T=\frac{g^2}{4\pi}\log\frac{\Lambda}{m_A}
 +\frac{g^2}{4}(C_A+C_I).
 \label{eq:Deltasigma}
\end{equation}
We therefore absorb the large-$R$ area and perimeter contributions into the physical parameters $\sigma_{\rm phys}$ and $T_{\rm phys}$ already appearing in \eqref{eq:diskNG}.  

After these local contributions are absorbed, the remaining finite-radius correction can be written in a cutoff-independent form as\footnote{Note that the apparent $R\log R$ term does not survive at large $R$. The logarithmic term is canceled by the $-\pi^{-1}\log(m_A R)$ contribution contained in the large-$m_A R$ expansion of $\mathcal I(m_A R)$.}
\begin{equation}
 E_{\rm FS}^{\rm ren}(R;m_A)
 =\frac{\pi g^2R}{2}
 \left[
 {\cal I}(m_AR)-\frac{m_AR}{2}
 +\frac{1}{\pi}\log(m_AR)-C_I
 \right].
 \label{eq:EFSren}
\end{equation}
The static disk energy is therefore
\begin{equation}
 E_{\rm disk}(R)
 =E_{\rm NG}^{\rm phys}(R)+E_{\rm FS}^{\rm ren}(R;m_A).
 \label{eq:diskren}
\end{equation}
This organization also makes the role of the thin-source approximation explicit: the cutoff-dependent core contribution is contained in $T_{\rm phys}$, whereas the residual finite-radius term in \eqref{eq:EFSren} is independent of the cutoff within the quadratic theory.

\begin{figure}[H]
 \centering
 \includegraphics[width=0.6\textwidth]{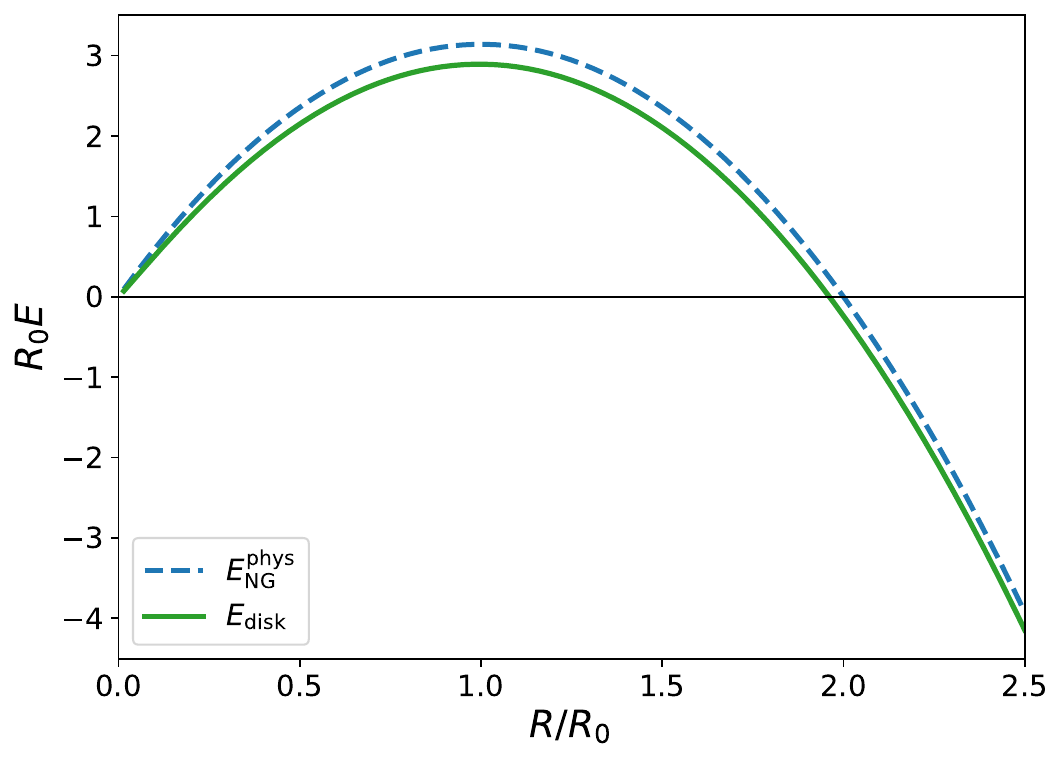}
 \caption{Renormalized static disk energy for the illustrative choice $T_{\rm phys}R_0^2=1$, $\sigma_{\rm phys}R_0^3=-1$, $m_A R_0=0.25$, and $gR_0=0.8$.  The horizontal axis is $R/R_0$, where $R_0=-T_{\rm phys}/\sigma_{\rm phys}$ is the radius at the maximum of the physical Nambu--Goto barrier, and the vertical axis is $R_0E$.  The blue dashed curve shows $R_0E_{\rm NG}^{\rm phys}$ without the finite-radius correction, while the green solid curve shows the full renormalized disk energy $R_0E_{\rm disk}=R_0(E_{\rm NG}^{\rm phys}+E_{\rm FS}^{\rm ren})$.}
 \label{fig:diskenergy}
\end{figure}

Figure~\ref{fig:diskenergy} illustrates the resulting correction for an illustrative choice of dimensionless parameters.  We take $m_AR_0=0.25$ and $gR_0=0.8$, for which the finite-radius correction is clearly visible near the barrier.

\section{A parameter-free quadratic prediction for the CSL}

\subsection{Wall gas and comparison with the exact CSL}

We now apply the quadratic effective field theory to the infinite CSL.  Consider an infinite periodic array of equally oriented parallel walls with lattice spacing $\ell$.  The interaction between two walls separated by a distance $n\ell$ is proportional to $e^{-mn\ell}$.  For a given reference wall, there are two walls at each distance $n\ell$, one on either side.  When computing the interaction energy per wall, however, each pair of walls must be counted only once.  The factor of two from the two sides is therefore canceled by the factor of $1/2$ that removes the double counting of wall pairs.  Thus, the interaction energy per wall and per unit area is
\begin{equation}
 \frac{E_{\rm int}^{(\mathrm{per\ wall})}}{A}
 =
 \sum_{n=1}^{\infty}
 C_{\rm dual}e^{-mn\ell}
 =
 \frac{C_{\rm dual}}{e^{m\ell}-1}.
 \label{eq:wallgasinteraction}
\end{equation}
Here $n$ labels the separation between a pair of walls in units of the lattice spacing $\ell$: $n=1$ corresponds to nearest-neighbor walls, $n=2$ to next-nearest-neighbor walls, and so on.
Using the physical wall tension from \eqref{eq:sigmaphys}, the energy per unit area assigned to each wall is the sum of its tension and the interaction energy in \eqref{eq:wallgasinteraction}.  Since there is one wall per distance $\ell$, the corresponding energy per unit volume is
\begin{equation}
 {\cal E}_{\rm dual}(\ell)
 =\frac{1}{\ell}
 \left[
 \sigma_{\rm phys}
 +\frac{C_{\rm dual}}{e^{m\ell}-1}
 \right].
 \label{eq:wallgasenergy}
\end{equation}
In terms of $x=m\ell$ and $\beta$ in \eqref{eq:sigmaphys}, we have $\sigma_{\rm phys}=8mf^2(1-\beta)$.  Minimizing \eqref{eq:wallgasenergy} with respect to the lattice spacing, or equivalently $x=m\ell$, gives
\begin{equation}
 -\frac{\sigma_{\rm phys}}{C_{\rm dual}}
 =\frac{e^x(1+x)-1}{(e^x-1)^2}.
 \label{eq:dualextremum}
\end{equation}
Substituting \eqref{eq:Cdual}, we obtain
\begin{equation}
 \beta_{\rm dual}(x)
 =1+\frac{\pi^2}{4}
 \frac{e^x(1+x)-1}{(e^x-1)^2},
 \qquad x=m\ell .
 \label{eq:betadual}
\end{equation}
With $C_{\rm dual}$ fixed by duality, the quadratic higher-form theory predicts the relation between the lattice spacing and the magnetic field for given $m$ and $f$. 

\begin{figure}[H]
 \centering
 \includegraphics[width=0.55\textwidth]{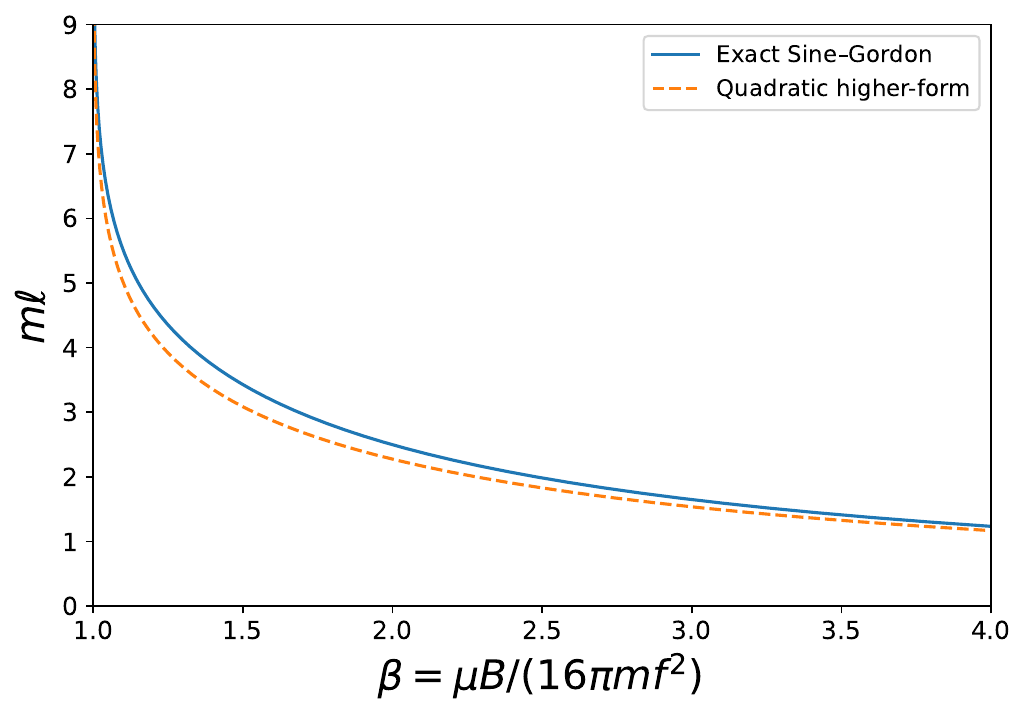}
 \caption{Dimensionless lattice spacing $m\ell$ as a function of $\beta=\mu B/(16\pi mf^2)$.  The solid curve is the exact Sine--Gordon CSL obtained from $\beta=E(k)/k$ and $m\ell=2kK(k)$.  The dashed curve is the quadratic higher-form prediction \eqref{eq:betadual}.  }
 \label{fig:spacing}
\end{figure}

For comparison, the exact Sine--Gordon result is given parametrically by
\begin{equation}
 \beta=\frac{E(k)}{k},
 \qquad
 x=2kK(k).
 \label{eq:exactparametric}
\end{equation}
Figure~\ref{fig:spacing} compares \eqref{eq:betadual} with \eqref{eq:exactparametric}.  Both theories predict $m\ell\to\infty$ as $\beta\to1$ and a monotonically decreasing spacing as the magnetic field is increased.  The quadratic dual curve remains close to the exact result even away from the asymptotically dilute regime.  The quadratic theory slightly underestimates the spacing, as expected if the repulsive interaction is somewhat weaker than in the exact soliton theory.

\subsection{Dilute limit and the order-one mismatch}

The agreement of Figure~\ref{fig:spacing} should not be confused with exact comparison of the two descriptions at quadratic order.  The dilute limit makes the difference particularly transparent.  Expanding the exact elliptic relations for $x=m\ell\gg1$ gives
\begin{equation}
 \beta_{\rm exact}-1
 \simeq4(1+x)e^{-x}.
 \label{eq:exactasymptotic}
\end{equation}
We can use this dilute-limit expansion to extract the asymptotic interaction coefficient between two well-separated walls.  Suppose that their interaction per unit area has the generic form
\begin{equation}
 \frac{V_{\rm pair}(\ell)}{A}=Ce^{-m\ell}.
\end{equation}
In the dilute limit, keeping the leading interaction between neighboring walls, the energy density takes the form
\begin{equation}
 {\cal E}(\ell)
 \simeq
 \frac{1}{\ell}
 \left[
  8mf^2(1-\beta)+Ce^{-m\ell}
 \right].
\end{equation}
Minimizing this energy with respect to the lattice spacing, or equivalently $x=m\ell$, gives
\begin{equation}
 \beta-1
 =\frac{C}{8mf^2}(1+x)e^{-x}.
 \label{eq:genericasymptotic}
\end{equation}
Comparing this expression with the exact dilute-limit result \eqref{eq:exactasymptotic} determines the exact asymptotic coefficient,
\begin{equation}
 C_{\rm exact}=32mf^2.
 \label{eq:Cexact}
\end{equation}
Therefore
\begin{equation}
 \frac{C_{\rm exact}}{C_{\rm dual}}
 =\frac{16}{\pi^2}
 \simeq1.621.
 \label{eq:ratio}
\end{equation}
The exponential range is reproduced correctly by the quadratic higher-form mediator, but its normalization differs by an order-one factor.  This is precisely what one expects from a truncation that retains the light mass scale and the topological normalization while discarding the detailed nonlinear soliton profile.  The mismatch motivates the full completion in the next section.

\section{Nonlinear multi-branch higher-form completion}

Nonlinear functions of a four-form field strength can encode general axion potentials, and an explicit three-form representation of a cosine potential has long been known~\cite{Dvali:2005}.  More recently, the nonlinear three-form description and its dual Kalb--Ramond formulation have been developed with particular attention to the associated global, multi-vacuum, and domain-wall/string structure~\cite{Anber:2024}.  Our aim in this section is to specialize this correspondence to the CSL and, in extending the description from an isolated wall to repeated $2\pi$ winding, to keep track of the branch information required to follow the scalar trajectory globally.

\subsection{Legendre dual of the cosine potential}

The quadratic $F_4^2$ theory does not retain the global periodic structure of the Sine--Gordon potential.  In Section~2, we started from the scalar theory written in terms of $\phi$ and derived its higher-form dual description.  Here it is useful to reverse the logic.  We start from a nonlinear higher-form theory and determine the conditions under which it reproduces the original scalar theory with potential $V(\phi)$.  In particular, the function $W({\cal F})$ will eventually be identified with the Legendre transform of $V(\phi)$.  Rather than imposing this relation from the outset, however, we leave $W({\cal F})$ unspecified and derive the condition it must satisfy from the equivalence of the two descriptions.

To this end, write
$
 F_4=dA_3={\cal F}\,*1,
$
where $*1$ denotes the oriented four-dimensional volume form, and consider the local higher-form action
\begin{equation}
 S_{\rm HF}
 =\frac{1}{8\pi^2f^2}\int H_3\wedge *H_3
 +\int d^4x\,W({\cal F}),
 \qquad
 H_3=dC_2-A_3.
 \label{eq:nonlinearHF}
\end{equation}

Variation of \eqref{eq:nonlinearHF} with respect to $A_3$ gives
\begin{equation}
 dW'({\cal F})
 =\frac{1}{4\pi^2f^2}*H_3,
 \label{eq:A3nonlinearEOM}
\end{equation}
where a prime denotes differentiation with respect to ${\cal F}$.  The duality relation allows us to reconstruct locally a scalar angle $\phi$ through
\begin{equation}
 *H_3=2\pi f^2 d\phi.
 \label{eq:dualityrelationnonlinear}
\end{equation}
Substituting this relation into \eqref{eq:A3nonlinearEOM} yields
\begin{equation}
 d\left[W'({\cal F})-\frac{\phi}{2\pi}\right]=0.
 \label{eq:Wprimeclosed}
\end{equation}
Thus, choosing the additive constant in $\phi$ appropriately,
\begin{equation}
 W'({\cal F})=\frac{\phi}{2\pi}.
 \label{eq:Wprimephi}
\end{equation}
To determine $W({\cal F})$, we must next relate ${\cal F}$ to the scalar potential $V(\phi)$.  This second condition follows from the modified Bianchi identity.  Since $H_3=dC_2-A_3$,
\begin{equation}
 dH_3=-F_4=-{\cal F}\,*1.
 \label{eq:Bianchi51}
\end{equation}
Together with \eqref{eq:dualityrelationnonlinear}, this gives the scalar equation in the form
\begin{equation}
 d*d\phi=-\frac{{\cal F}}{2\pi f^2}\,*1.
 \label{eq:scalarfromBianchi51}
\end{equation}
For comparison, a scalar theory with potential $V(\phi)$ obeys
\begin{equation}
 d*d\phi=-\frac{1}{f^2}\frac{dV}{d\phi}\,*1.
 \label{eq:scalarEOMV}
\end{equation}
Therefore the higher-form theory reproduces the same scalar dynamics only if
\begin{equation}
 \frac{{\cal F}}{2\pi}=\frac{dV}{d\phi}.
 \label{eq:FfromVgeneral}
\end{equation}
Equations~\eqref{eq:Wprimephi} and \eqref{eq:FfromVgeneral} now determine the function $W({\cal F})$.  From Eq.~\eqref{eq:Wprimephi}, the variation of $W$ along the scalar trajectory is
\begin{equation}
 dW
 =W'({\cal F})\,d{\cal F}
 =\frac{\phi}{2\pi}\,d{\cal F}.
\end{equation}
On the other hand, Eq.~\eqref{eq:FfromVgeneral} implies
\begin{equation}
 dV(\phi)
 =V'(\phi)\,d\phi
 =\frac{{\cal F}}{2\pi}\,d\phi.
\end{equation}
The differential of the product $\phi{\cal F}/(2\pi)$ can therefore be written as
\begin{equation}
 d\left(\frac{\phi{\cal F}}{2\pi}\right)
 =\frac{\phi}{2\pi}\,d{\cal F}
 +\frac{{\cal F}}{2\pi}\,d\phi
 =dW+dV.
\end{equation}
It follows that
\begin{equation}
 dW
 =d\left[\frac{\phi{\cal F}}{2\pi}-V(\phi)\right],
\end{equation}
and hence, up to an additive constant,
\begin{equation}
 W({\cal F})
 =\frac{\phi{\cal F}}{2\pi}-V(\phi),
 \qquad
 \frac{{\cal F}}{2\pi}=V'(\phi).
 \label{eq:LegendreWderived}
\end{equation}
Thus equivalence with the scalar theory requires $W({\cal F})$ to be the Legendre transform of $V(\phi)$, with conjugate variable ${\cal F}/(2\pi)=V'(\phi)$.  The Legendre-transform relation is therefore not imposed at the outset, but follows from requiring the nonlinear higher-form theory to reproduce the scalar dynamics.

For the Sine--Gordon potential
$
 V(\phi)=f^2m^2(1-\cos\phi),
 $
we obtain $ {\cal F}=2\pi f^2m^2\sin\phi.
$
It is convenient to define
\begin{equation}
 y\equiv\frac{{\cal F}}{2\pi f^2m^2}=\sin\phi.
 \label{eq:ydef}
\end{equation}
The Legendre transform has the parametric form
\begin{equation}
 W=f^2m^2\left[\phi\sin\phi-(1-\cos\phi)\right].
 \label{eq:Wparam}
\end{equation}
On the principal branch, $\phi=\arcsin y$, this becomes
\begin{equation}
 W_0(y)
 =f^2m^2\left[
 y\arcsin y-1+\sqrt{1-y^2}
 \right].
 \label{eq:Wprincipal}
\end{equation}
At small field,
\begin{equation}
 W_0
 =f^2m^2\left[
 \frac{y^2}{2}+\frac{y^4}{24}+\frac{y^6}{80}+O(y^8)
 \right].
 \label{eq:Wexpansion}
\end{equation}
The leading term is precisely the quadratic four-form theory.  Thus the effective field theory of the previous sections is the first term in a systematic nonlinear completion rather than an unrelated phenomenological model.

\subsection{Why the dual theory is multi-branch}

The map $y=\sin\phi$ is not one-to-one.  Global branch structure is a familiar feature of axion--three-form systems \cite{Anber:2024}.  For the CSL the relevant question is how this information is followed continuously through each $2\pi$ winding.  For a given $y$ one must retain the branches
\begin{align}
 \phi_{n,+}&=2\pi n+\arcsin y,
 \\
 \phi_{n,-}&=(2n+1)\pi-\arcsin y,
 \label{eq:branches}
\end{align}
with $n\in\mathbb Z$.  Substituting these inverse branches into the Legendre transform \eqref{eq:Wparam}, and using $y={\cal F}/(2\pi f^2m^2)$, defines the corresponding branches of $W({\cal F})$:
\begin{align}
 W_{n,+}({\cal F})
 &=f^2m^2\left[
 y\left(2\pi n+\arcsin y\right)-1+\sqrt{1-y^2}
 \right],
 \nonumber\\
 W_{n,-}({\cal F})
 &=f^2m^2\left[
 y\left((2n+1)\pi-\arcsin y\right)-1-\sqrt{1-y^2}
 \right].
 \label{eq:Wbranches}
\end{align}
For each branch the dimensionless four-form variable lies in the range $-1\leq y\leq1$.
Consequently the exact function $W({\cal F})$ is not single-valued, but consists of a collection of branches $W_{n,\pm}({\cal F})$, denoted collectively by $W_{\rm multi}({\cal F})$.

This branch structure is not an additional microscopic degree of freedom.  It keeps track of information that would otherwise be lost when the compact scalar is mapped to ${\cal F}\propto\sin\phi$.  To display the branch structure explicitly, we divide the continuous trajectory of $\phi$ at the turning points of the map $y=\sin\phi$, where $y=\pm1$. Each branch then covers an interval of length $\pi$ in $\phi$, and the branch assignment can be written as
\begin{equation}
 W_{\rm multi}({\cal F})=
 \begin{cases}
  W_{0,+}({\cal F}),
  & (-{\pi}/{2}\leq\phi\leq {\pi}/{2}),
  \\[4pt]
  W_{0,-}({\cal F}),
  & ({\pi}/{2}\leq\phi\leq {3\pi}/{2}),
  \\[4pt]
  W_{1,+}({\cal F}),
  & ({3\pi}/{2}\leq\phi\leq {5\pi}/{2}),
  \\[4pt]
&  \vdots 
 \end{cases}
 \label{eq:wallbranchpath}
\end{equation}
The neighboring branches are connected continuously at the turning points $y=\pm1$:
\begin{equation}
 \left.W_{n,+}({\cal F})\right|_{y=1}
 =
 \left.W_{n,-}({\cal F})\right|_{y=1},
 \qquad
 \left.W_{n,-}({\cal F})\right|_{y=-1}
 =
 \left.W_{n+1,+}({\cal F})\right|_{y=-1}.
 \label{eq:Wbranchjoins}
\end{equation}
Thus the branch flow proceeds as
\begin{equation}
 W_{n,+}
 \xrightarrow{\,y=1\,}
 W_{n,-}
 \xrightarrow{\,y=-1\,}
 W_{n+1,+},
 \label{eq:Wbranchflow}
\end{equation}
A fundamental wall, for which $\phi$ varies from $0$ to $2\pi$, therefore follows the branch sequence $W_{0,+}\to W_{0,-}\to W_{1,+}$, with the branch changes occurring at $y=1$ and $y=-1$, respectively.
Although the local four-form variable returns to $y=0$ at the end of the wall, the trajectory has arrived on the neighboring sheet $W_{1,+}$ rather than returning to its initial sheet $W_{0,+}$.  Continuing to larger $\phi$ repeats the same pattern with the branch label shifted by one.

\subsection{Full higher-form action}

Having determined $W_{\rm multi}({\cal F})$, the nonlinear higher-form description can be written entirely in terms of $C_2$ and $A_3$.  The full dual action is
\begin{equation}
 S_{\rm full}[C_2,A_3]
 =\frac{1}{8\pi^2f^2}\int H_3\wedge *H_3
 +\int d^4x\,W_{\rm multi}({\cal F})
 +S_{\mu B}^{\rm dual},
 \label{eq:fullaction}
\end{equation}
At this stage no explicit Nambu--Goto wall or string source is added: the smooth wall and the CSL are classical solutions of the nonlinear higher-form fields themselves.  The explicit source description of Sections~2--4 is the corresponding thin-defect effective field theory obtained after coarse graining.

The anomaly term can also be written directly in higher-form variables.  Introduce a background baryon gauge field $A_B=\mu\,dt$ and denote the electromagnetic field strength by $F_{\rm em}$.  We use the convention
\begin{equation}
 B^i=\frac{1}{2}\epsilon^{ijk}(F_{\rm em})_{jk},
 \qquad
 \epsilon^{123}=+1,
 \label{eq:Bconvention}
\end{equation}
so that for $\mathbf B=B\hat{\mathbf z}$ one has $F_{\rm em}=B\,dx\wedge dy$.  The scalar term is
\begin{equation}
 S_{\mu B}
 =\frac{1}{4\pi^2}\int A_B\wedge d\phi\wedge F_{\rm em}.
 \label{eq:anomalyscalar}
\end{equation}
For $\mathbf B=B\hat{\mathbf z}$ and $\phi=\phi(z)$, this gives
$A_B\wedge d\phi\wedge F_{\rm em}=\mu B\phi'(z)d^4x$, in agreement with the convention used in Eq.~\eqref{eq:SGaction}.
Using \eqref{eq:dualityrelation}, it becomes
\begin{equation}
 S_{\mu B}^{\rm dual}
 =\frac{1}{8\pi^3f^2}
 \int A_B\wedge *H_3\wedge F_{\rm em}.
 \label{eq:anomalydual}
\end{equation}
For a fundamental wall this reproduces the energy shift
\begin{equation}
 \Delta\sigma_{\mu B}=-\frac{\mu B}{2\pi},
 \label{eq:wallshiftdual}
\end{equation}
and hence the same critical condition $\mu B=16\pi mf^2$ as the scalar description.

\subsection{Reconstructing the Sine--Gordon equation from higher-form variables}

We now use $W_{\rm multi}$ to check explicitly that the resulting higher-form theory reproduces the Sine--Gordon equation.  In deriving the local bulk equations below, we omit the anomaly term $S_{\mu B}^{\rm dual}$.  In the scalar description this term is a total derivative: it changes the energy of a winding sector but does not modify the local Sine--Gordon equation.  Its contribution to the energy will be restored below.  The independent fields in the remaining bulk action are $C_2$ and $A_3$, with ${\cal F}$ defined through $F_4=dA_3={\cal F}\,*1$.  The $A_3$ equation of motion is then
\begin{equation}
 dW'_{\rm multi}({\cal F})
 =\frac{1}{4\pi^2f^2}*H_3.
 \label{eq:A3eomfull}
\end{equation}
It is convenient to introduce the quantity
\begin{equation}
 u({\cal F})\equiv2\pi W'_{\rm multi}({\cal F}).
 \label{eq:udef}
\end{equation}
To distinguish this quantity, reconstructed entirely from the higher-form variables, from the scalar field $\phi$ used in Sec.~2, we denote it by $u$ for the moment.  In terms of $u$, the $A_3$ equation \eqref{eq:A3eomfull} becomes
\begin{equation}
 *H_3=+2\pi f^2du.
 \label{eq:Hdu}
\end{equation}
The second relation needed to determine the dynamics follows from the specific nonlinear function $W_{\rm multi}$ constructed above.  Since it is the Legendre transform associated with the Sine--Gordon potential, its derivative gives the relation
\begin{equation}
 \frac{{\cal F}}{2\pi f^2m^2}=\sin u,
 \label{eq:constitutiveu}
\end{equation}
Finally, the definition $H_3=dC_2-A_3$ gives the modified Bianchi identity
\begin{equation}
 dH_3=-F_4=-{\cal F}\,*1.
 \label{eq:Bianchi}
\end{equation}
Thus \eqref{eq:Hdu}, \eqref{eq:constitutiveu}, and \eqref{eq:Bianchi} form a closed set of relations within the higher-form description.  In four-dimensional Lorentzian spacetime with our mostly-minus convention, $**=+1$ on one- and three-forms.  Applying the Hodge star to Eq.~\eqref{eq:Hdu} therefore gives
\begin{equation}
 H_3=2\pi f^2*du,
 \label{eq:Hdualityu}
\end{equation}
in the same form as the duality relation \eqref{eq:dualityrelation}.  Taking an exterior derivative of Eq.~\eqref{eq:Hdualityu} and using Eqs.~\eqref{eq:constitutiveu} and \eqref{eq:Bianchi} gives
\begin{equation}
 d*du=-m^2\sin u\,*1.
 \label{eq:SGfromHF}
\end{equation}
For a static configuration depending on $z$ only,
\begin{equation}
 u''(z)=m^2\sin u(z).
 \label{eq:SGufinal}
\end{equation}
The quantity $u$ was introduced entirely within the higher-form theory through $u({\cal F})=2\pi W'_{\rm multi}({\cal F})$. Equation~\eqref{eq:SGufinal}, derived solely from the higher-form equations, is precisely the Sine--Gordon equation obeyed by the scalar angle $\phi$ in the original scalar description. We can therefore identify the reconstructed variable $u$ with $\phi$. In this sense the nonlinear $C_2$--$A_3$ theory reproduces the scalar dynamics rather than importing the known scalar solution into the dual equations.

\subsection{Exact CSL and branch flow}

\eqref{eq:SGufinal} immediately gives the exact periodic solution
\begin{equation}
 u(z)=\pi+2\,\mathrm{am}\left(\frac{mz}{k},k\right),
 \qquad
 \ell=\frac{2kK(k)}{m}.
 \label{eq:uexact}
\end{equation}
The higher-form fields themselves are periodic:
\begin{equation}
 \frac{{\cal F}(z)}{2\pi f^2m^2}
 =\sin u(z)
 =-2\,\mathrm{sn}\!\left(\frac{mz}{k},k\right)\,\mathrm{cn}\!\left(\frac{mz}{k},k\right),
 \label{eq:FCSL}
\end{equation}
and \eqref{eq:Hdu} gives a component of $H_3$ proportional to $\mathrm{dn}\!\left(\frac{mz}{k},k\right)$.  The scalar angle, however, changes by $2\pi$ over one period:
\begin{equation}
 u(z+\ell)=u(z)+2\pi,
 \qquad
 {\cal F}(z+\ell)={\cal F}(z).
 \label{eq:periodicbranch}
\end{equation}
The missing information is carried by the branch label of $W_{\rm multi}$.  A single wall interpolates between two neighboring branches, $n$ and $n+1$, while ${\cal F}$ remains zero on both sides, and the CSL is a spatially periodic repetition of this branch change.

The equivalence also extends to the energy.  Here $E_{\rm per}$ denotes the energy per unit transverse area integrated over one CSL period of length $\ell$.  Accordingly, the corresponding energy per unit volume is $E_{\rm per}/\ell$, which is the quantity analogous to ${\cal E}_{\rm dual}$ used in Section~4.  Using the duality relation together with the Legendre relation between $W({\cal F})$ and $V(\phi)$, the higher-form energy evaluated on the reconstructed scalar configuration reduces to the scalar Sine--Gordon energy.  Excluding the anomaly contribution, we obtain
\begin{equation}
 E_{\rm per}^{(0)}
 =\frac{4mf^2}{k}
 \left[2E(k)-(1-k^2)K(k)\right].
 \label{eq:energyperperiod}
\end{equation}
The anomaly term contributes $-\mu B/(2\pi)$ for each $2\pi$ winding, exactly as in the scalar description.  The total energy per period is therefore the same as that obtained in Section~2.  Minimizing it with respect to the CSL modulus $k$, or equivalently the lattice spacing, reproduces \eqref{eq:exactcondition}.  Thus the nonlinear higher-form theory reproduces both the Sine--Gordon equation and the exact energetics that determine the CSL spacing.

\section{Dynamics of the nonlinear higher-form CSL}

The nonlinear completion does more than reconstruct the static periodic solution.  It also retains the dynamical information carried by the full soliton profile and by collective displacements of the lattice.  We now use the nonlinear higher-form theory to recover two physical properties of the CSL: the exact asymptotic wall--wall force and the gapless phonon mode.

\subsection{Exact dilute wall interaction from the nonlinear higher-form stress tensor}
\label{subsec:fullforce}

The exact CSL asymptotics in Section~4 imply $C_{\rm exact}=32mf^2$.  We now recover the same coefficient directly from the nonlinear higher-form theory.  The result also explains why the quadratic theory gets the range of the force right but misses its normalization.

The force between two parallel walls can be obtained from the component of the stress tensor normal to the walls. For walls lying in the $xy$ plane, this component is $T_{zz}$, whose value gives the pressure acting on the walls. We therefore evaluate $T_{zz}$ in the nonlinear higher-form theory. For a static planar configuration $u=u(z)$, metric variation gives
\begin{equation}
 T_{zz}
 =\frac{1}{8\pi^2f^2}\left((*H_3)_z\right)^2
 -\left[ {\cal F}W'({\cal F})-W({\cal F})\right].
 \label{eq:TzzHF}
\end{equation}
A derivation directly from the nonlinear higher-form action is given in Appendix~\ref{app:stress}.
Using $*H_3=2\pi f^2du$ and ${\cal F}W'({\cal F})-W({\cal F})=f^2m^2(1-\cos u)$, this becomes
\begin{equation}
 T_{zz}
 =\frac{f^2}{2}(u')^2-f^2m^2(1-\cos u).
 \label{eq:Tzzu}
\end{equation}
Equation~\eqref{eq:SGufinal} then shows that $T_{zz}$ is constant in the region between the walls, namely $dT_{zz}/dz=0$. We can therefore evaluate it at any convenient point in this region. For two widely separated walls, the field between them is determined by the exponentially small tails of the individual wall profiles. We first determine this tail from the exact single-wall solution and then use it to evaluate $T_{zz}$ between the two walls.

Consider a single wall for which the reconstructed scalar $u$ varies from $0$ to $2\pi$. In the higher-form description, this trajectory cannot remain on a single branch of $W_{\rm multi}({\cal F})$. Since ${\cal F}/(2\pi f^2m^2)=\sin u$, the wall follows the sequence $W_{0,+}\to W_{0,-}\to W_{1,+}$, with the branch changes occurring at $u=\pi/2$ and $u=3\pi/2$, where ${\cal F}/(2\pi f^2m^2)=+1$ and $-1$, respectively. Across these branch changes, the reconstructed field $u$ remains continuous and obeys the Sine--Gordon equation \eqref{eq:SGufinal}. Its single-wall solution is therefore
\begin{equation}
 u_{\rm W}(z)=4\arctan e^{m(z-z_0)},
\end{equation}
whose right tail is
\begin{equation}
 2\pi-u_{\rm W}(z)
 =4e^{-m(z-z_0)}+O(e^{-3m(z-z_0)}).
 \label{eq:fulltail}
\end{equation}
The exponential decay rate $m$ follows from the linearized equation around the vacuum, whereas the coefficient $4$ is fixed by the full nonlinear wall solution. This tail amplitude will determine the leading interaction between two well-separated walls.  For two successive walls centered at $z=\pm\ell/2$, the field in the middle region at $m\ell\gg1$ is
\begin{equation}
 u(z)\simeq2\pi
 -4e^{-m(z+\ell/2)}
 +4e^{m(z-\ell/2)}.
 \label{eq:twowalltail}
\end{equation}
At the midpoint,
\begin{equation}
 u(0)=2\pi+O(e^{-m\ell}),
 \qquad
 u'(0)=8m e^{-m\ell/2}+O(e^{-3m\ell/2}).
\end{equation}
Since $u(0)=2\pi$ to leading order, the potential term in Eq.~\eqref{eq:Tzzu} vanishes at the midpoint. The pressure between the walls is therefore determined entirely by the derivative term at this order.  The repulsive pressure is therefore
\begin{equation}
 \frac{F_{\rm rep}}{A}
 =T_{zz}(0)
 =32f^2m^2e^{-m\ell}+O(e^{-2m\ell}).
 \label{eq:fullpressure}
\end{equation}
Writing $V_{\rm int}/A=C_{\rm full}e^{-m\ell}+\cdots$ gives
\begin{equation}
 C_{\rm full}=32mf^2=C_{\rm exact}.
 \label{eq:Cfull}
\end{equation}
Thus the nonlinear higher-form theory directly restores the exact dilute coefficient.  The exponential $e^{-m\ell}$ is controlled by the linearized massive mode and is already reproduced by the quadratic dual theory, whereas the prefactor depends on the nonlinear tail amplitude in \eqref{eq:fulltail}.  This identifies the origin of the factor $16/\pi^2$ in \eqref{eq:ratio}.

\subsection{CSL phonon as a displacement of the higher-form background}
\label{subsec:phonon}

A useful test of the full formulation is whether it also captures the gapless collective mode associated with the broken continuous translation symmetry of the CSL.  In the scalar formulation this phonon is known to obey an $n=1$ Lam\'e spectral problem \cite{Brauner:2016pko}.  In the higher-form formulation, the same mode can be described as a fluctuation of the periodic higher-form background.

We first relate the fluctuation variable used below to the higher-form fields.  The CSL background is represented equivalently by the periodic profiles ${\cal F}_0(z)$ and $H_{3,0}(z)$.  Let $u_0(z)$ denote the exact CSL background in Eq.~\eqref{eq:uexact}.  Using the duality relations $*H_3=2\pi f^2du$ and ${\cal F}=2\pi f^2m^2\sin u$, a small fluctuation $u=u_0+\eta$ induces
\begin{equation}
 \delta(*H_3)=2\pi f^2d\eta,
 \qquad
 \delta{\cal F}=2\pi f^2m^2\cos u_0\,\eta.
\end{equation}

Linearizing \eqref{eq:SGfromHF} gives
\begin{equation}
 \left[\partial_\mu\partial^\mu+m^2\cos u_0(z)\right]\eta=0.
 \label{eq:phononlin}
\end{equation}
To separate the dependence along the wall from that in the normal direction, we write
\begin{equation}
 \eta=e^{-i\omega t+ip_xx+ip_yy}\psi(z).
\end{equation}
Using $\partial_\mu\partial^\mu=\partial_t^2-\partial_x^2-\partial_y^2-\partial_z^2$, the linearized equation first reduces to the one-dimensional eigenvalue problem
\begin{equation}
 \left[-\frac{d^2}{dz^2}+m^2\cos u_0(z)\right]\psi(z)
 =\left(\omega^2-p_x^2-p_y^2\right)\psi(z).
\end{equation}
For the exact CSL background, it is convenient to introduce the dimensionless coordinate $s\equiv mz/k$.  In terms of this coordinate,
\[
 \cos u_0(z)=2\operatorname{sn}^2(s,k)-1,
 \qquad
 \frac{d}{dz}=\frac{m}{k}\frac{d}{ds}.
\]
Substituting these relations into the equation above, the fluctuation problem reduces to the standard $n=1$ Lam\'e equation,
\begin{equation}
 \left[
  -\frac{d^2}{ds^2}
  +2k^2\operatorname{sn}^2(s,k)
 \right]\psi
 =\lambda\psi,
 \qquad
 \lambda\equiv
 k^2\left[
  1+\frac{\omega^2-p_x^2-p_y^2}{m^2}
 \right].
\end{equation}
Differentiating the background equation gives the translational mode
\begin{equation}
 \psi_0(z)=u_0'(z)
 =\frac{2m}{k}\operatorname{dn}(s,k).
 \label{eq:phononzero}
\end{equation}
Using the Lam\'e identity
\begin{equation}
 \left[-\frac{d^2}{ds^2}+2k^2\operatorname{sn}^2(s,k)\right]\operatorname{dn}(s,k)
 =k^2\operatorname{dn}(s,k),
\end{equation}
we find $\lambda=k^2$.  Comparing this with the definition of $\lambda$ above gives $\omega^2-p_x^2-p_y^2=0$.  In particular, at zero transverse momentum, $p_x=p_y=0$, one has $\omega=0$, which is why $u_0'(z)$ is the translational zero mode.

This is directly related to the usual Nambu--Goto interpretation of a wall zero mode.  For one isolated wall, in static gauge one may write its embedding as
\begin{equation}
 X^\mu(\xi^a)=\bigl(t,x,y,Z(t,x,y)\bigr),
\end{equation}
so that $Z$ is the massless fluctuation of the wall position in the normal direction.  Microscopically, $u_{\rm W}(z-Z)\simeq u_{\rm W}(z)-Z u_{\rm W}'(z)$, which is precisely the translational zero mode.  For a uniform displacement of the CSL in the $z$ direction, the translational zero mode can be written as
\begin{equation}
 \eta(t,x,y,z)=-\xi(t,x,y)u_0'(z).
 \label{eq:phonondisplacement}
\end{equation}
Here $\xi(t,x,y)$ describes the displacement of the CSL as a whole, while the $z$ dependence of the zero mode is contained in $u_0'(z)$.  To describe a phonon propagating along the lattice, this displacement can be allowed to vary slowly also in the $z$ direction, $\xi=\xi(t,x,y,z)$.  Equivalently, the discrete displacements of the individual walls, $z_n=n\ell+\xi_n$, become a slowly varying collective displacement field in the long-wavelength limit.

For a mode with frequency $\omega$ and transverse momenta $p_x$ and $p_y$, we write
\[
 \xi(t,x,y,z)=e^{-i\omega t+ip_xx+ip_yy}\xi(z),
\]
so that
\[
 \eta(t,x,y,z)=-e^{-i\omega t+ip_xx+ip_yy}u_0'(z)\xi(z).
\]
Thus, in the notation used below, $\psi(z)=-u_0'(z)\xi(z)$.  The fluctuation equation can be written as
\begin{equation}
 \left[-\partial_z^2+m^2\cos u_0(z)\right]\psi(z)
 =\varepsilon\psi(z),
 \qquad \varepsilon\equiv\omega^2-p_x^2-p_y^2.
\end{equation}
Defining $\rho=\psi_0^2$ gives
\begin{equation}
 -\frac{d}{dz}\left(\rho\frac{d\xi}{dz}\right)
 =\varepsilon\rho\xi.
\end{equation}
For a mode with Bloch momentum $p_z$, we impose the Bloch condition $\xi(z+\ell)=e^{ip_z\ell}\xi(z)$.  For $p_z\ell\ll1$, the lowest band is
\begin{equation}
 \varepsilon=c_z^2p_z^2+O(p_z^4),
 \qquad
 c_z^2=\frac{\ell^2}
 {\left(\int_0^\ell dz\,\rho\right)
  \left(\int_0^\ell dz\,\rho^{-1}\right)}.
 \label{eq:czgeneral}
\end{equation}
A derivation of Eq.~\eqref{eq:czgeneral} is given in Appendix~\ref{app:phononlong}.
Using $\rho=(2m/k)^2\operatorname{dn}^2(s,k)$ together with
\begin{equation}
 \int_0^{2K}ds\,\operatorname{dn}^2(s,k)=2E(k),
 \qquad
 \int_0^{2K}\frac{ds}{\operatorname{dn}^2(s,k)}
 =\frac{2E(k)}{1-k^2},
\end{equation}
one obtains
\begin{equation}
 c_z^2=(1-k^2)\left[\frac{K(k)}{E(k)}\right]^2,
\end{equation}
and hence
\begin{equation}
 \omega^2
 =p_x^2+p_y^2
 +(1-k^2)\left[\frac{K(k)}{E(k)}\right]^2p_z^2
 +O(p_z^4).
 \label{eq:phonondispersion}
\end{equation}
This agrees with the known CSL phonon dispersion \cite{Brauner:2016pko}.  In particular the longitudinal velocity vanishes as $k\to1$ as the CSL first appears and approaches unity as $k\to0$.

The fluctuation also sharpens the role of the branch variable.  Since $\delta{\cal F}\propto\cos u_0\,\eta$, ${\cal F}$ alone cannot be used globally as a nonsingular phonon coordinate: at points where $\cos u_0=0$, the map from $\eta$ to $\delta{\cal F}$ degenerates.  The branch information, or equivalently the smooth reconstructed variable $u$, remains essential even for the low-energy collective dynamics.  The branch-flow picture therefore extends beyond the static CSL.

\section{Discussion and outlook}

We have applied a dual framework to the chiral soliton lattice, first using the quadratic higher-form theory and then extending it to the nonlinear theory that reproduces the full Sine--Gordon dynamics.

At the quadratic level, the wall and string are treated as infinitely thin sources coupled to the higher-form fields.  The resulting theory predicts the CSL spacing without an additional fitting parameter and reproduces the exact result reasonably well.  In the dilute limit, however, the wall--wall interaction coefficient differs from the exact result by a factor $16/\pi^2$, quantifying the limitation of treating the wall as an infinitely thin source in the quadratic theory.

The nonlinear higher-form theory restores the information missing from this approximation.  By following the appropriate branches of $W_{n,\pm}({\cal F})$, repeated scalar winding is retained even though the four-form field strength returns to the same value after each period.  This construction reproduces the exact Sine--Gordon dynamics and CSL solution, and yields the exact dilute interaction coefficient $C_{\rm full}=32mf^2$.

The same full higher-form theory also captures the collective low-energy dynamics of the lattice.  For a single wall, the translational zero mode describes a displacement of the wall in the direction normal to its surface.  In a lattice the individual wall displacements combine into a long-wavelength field $\xi(t,x,y,z)$, which is the CSL phonon.  Linearizing the higher-form equations about the periodic background gives the $n=1$ Lam\'e problem and reproduces
\begin{equation}
 \omega^2=p_x^2+p_y^2+(1-k^2)\left[\frac{K(k)}{E(k)}\right]^2p_z^2+O(p_z^4),
 \label{eq:discussionphonon}
\end{equation}
in agreement with the scalar analysis \cite{Brauner:2016pko}.  In higher-form variables this mode is simply a local displacement of the periodic $H_3$ and $F_4$ profiles.  Thus the branch-flow description extends from the static CSL to its gapless collective dynamics.

One possible application of the higher-form framework developed here is to the formation and decay of finite wall--string configurations.  The description of an axion string as the boundary of an open domain-wall membrane already appears in earlier effective descriptions of wall--string systems \cite{Townsend:1993}.  In the CSL context a wall disk bounded by a string loop provides the relevant collective description for quantum nucleation \cite{Higaki:2022qwt,Eto:2022lcs}.  The present formulation provides a gauge-invariant framework in which the wall, its boundary string, and the effects of the massive higher-form fields can be treated together.  A detailed semiclassical calculation of nucleation rates, including the relation between the leading Nambu--Goto limit and corrections from the massive sector, will be discussed separately.

A second direction is to generalize the vacuum structure.  QCD at nonzero $\theta$ can contain several physically relevant vacuum sheets and elementary wall transitions within one $2\pi$ excursion.  In such systems the number of elementary walls should not be identified naively with the integer multiplying the St\"uckelberg coupling: a single unit of winding may decompose into several wall transitions while the total higher-form charge remains fixed.  This leads to nontrivial string--wall junction rules that are naturally constrained by higher-form gauge invariance.  We leave this richer multi-sheet problem for a separate study.

Finally, generalizations to St\"uckelberg couplings with integer coefficients larger than one are potentially relevant to axion string--wall systems, where discrete higher-form structures and domain-wall number become important \cite{Hidaka:2019threeform,BerasaluceGonzalez:2012zn,Kibble:1982dd,Vilenkin:1982ks,Preskill:1992ck,Svrcek:2006yi,Marsh:2015xka}.  In related string compactifications, integer BF couplings lead to discrete gauge structures and to configurations in which multiple domain walls can be attached to a string \cite{BerasaluceGonzalez:2012zn}.  The precise relation between a St\"uckelberg integer, an anomaly coefficient, and the number of elementary walls attached to a string depends on the microscopic theory and should be established rather than assumed.  The CSL studied here avoids these additional complications while retaining the nontrivial winding and wall dynamics that are central to the lattice.

\section*{Acknowledgments}

This work is supported by Grant-in-Aid for Scientific Research from the Ministry of Education, Culture, Sports, Science and Technology, Japan (JP24K07022). The author used ChatGPT (OpenAI, GPT-5.6) as an aid for scientific discussion, numerical exploration, code development, and manuscript preparation. The author reviewed and verified the resulting calculations and interpretations and takes full responsibility for the content.

\appendix

\section{Stress tensor of the nonlinear higher-form theory}
\label{app:stress}

In this appendix we derive the component of the stress tensor used in Eq.~\eqref{eq:TzzHF}. It is useful to perform the variation before specializing to flat spacetime. The covariant form of the nonlinear bulk action relevant for this calculation is
\begin{equation}
 S_{\rm bulk}
 =\frac{1}{8\pi^2f^2}\int d^4x\,\sqrt{-g}\,\frac{1}{3!}H_{\mu\nu\rho}H^{\mu\nu\rho}
 +\int d^4x\,\sqrt{-g}\,W({\cal F}),
 \label{eq:appbulk}
\end{equation}
where $\epsilon_{\mu\nu\rho\sigma}$ is the Levi--Civita tensor,
\begin{equation}
 F_{\mu\nu\rho\sigma}={\cal F}\,\epsilon_{\mu\nu\rho\sigma}.
 \label{eq:appFdef}
\end{equation}
We choose the orientation such that $\epsilon_{0123}=+\sqrt{-g}$. With the metric signature $(+,-,-,-)$, one then has $\epsilon^{0123}=-1/\sqrt{-g}$. We define
\begin{equation}
 T_{\mu\nu}=\frac{2}{\sqrt{-g}}\frac{\delta S_{\rm bulk}}{\delta g^{\mu\nu}}.
 \label{eq:appTdef}
\end{equation}

For the $H_3$ term, using
\begin{equation}
 \delta\sqrt{-g}=-\frac12\sqrt{-g}\,g_{\mu\nu}\delta g^{\mu\nu}
 \label{eq:appdetvar}
\end{equation}
together with the variation of the inverse metrics that raise the indices of $H_{\mu\nu\rho}$ gives
\begin{equation}
 T_{\mu\nu}^{(H)}
 =\frac{1}{8\pi^2f^2}
 \left[
 H_{\mu\alpha\beta}H_{\nu}{}^{\alpha\beta}
 -\frac{1}{3!}g_{\mu\nu}H_{\alpha\beta\gamma}H^{\alpha\beta\gamma}
 \right].
 \label{eq:appTH}
\end{equation}

The variation of the $W({\cal F})$ term requires the metric dependence of ${\cal F}$. Keeping the differential form $F_4$ fixed, Eq.~\eqref{eq:appFdef} implies
\begin{equation}
 \delta{\cal F}=\frac12{\cal F}\,g_{\mu\nu}\delta g^{\mu\nu}.
 \label{eq:appFvar}
\end{equation}
Therefore
\begin{equation}
 \delta\!\left(\sqrt{-g}W({\cal F})\right)
 =\frac12\sqrt{-g}\,g_{\mu\nu}
 \left[{\cal F}W'({\cal F})-W({\cal F})\right]\delta g^{\mu\nu},
\end{equation}
and hence
\begin{equation}
 T_{\mu\nu}^{(W)}
 =g_{\mu\nu}\left[{\cal F}W'({\cal F})-W({\cal F})\right].
 \label{eq:appTW}
\end{equation}

We now specialize to a static planar configuration depending only on $z$. In this case $*H_3$ has only a $z$ component. Equivalently, $H_3$ has components only along the $(t,x,y)$ directions, so that $H_{z\alpha\beta}=0$. With the conventions above, the covariant identity is
\begin{equation}
 \frac{1}{3!}H_{\alpha\beta\gamma}H^{\alpha\beta\gamma}
 =-(*H_3)_\mu(*H_3)^\mu.
 \label{eq:appHdualidentity}
\end{equation}
For the present planar configuration, $(*H_3)_\mu$ has only a $z$ component and $g^{zz}=-1$, so this becomes
\begin{equation}
 \frac{1}{3!}H_{\alpha\beta\gamma}H^{\alpha\beta\gamma}
 =\left((*H_3)_z\right)^2.
 \label{eq:appHdual}
\end{equation}
Since $g_{zz}=-1$, Eqs.~\eqref{eq:appTH} and \eqref{eq:appTW} give
\begin{equation}
 T_{zz}
 =\frac{1}{8\pi^2f^2}\left((*H_3)_z\right)^2
 -\left[{\cal F}W'({\cal F})-W({\cal F})\right],
 \label{eq:appTzz}
\end{equation}
which is Eq.~\eqref{eq:TzzHF}.

Finally, using
\begin{equation}
 *H_3=2\pi f^2du,
 \qquad
 {\cal F}W'({\cal F})-W({\cal F})=f^2m^2(1-\cos u),
\end{equation}
we recover
\begin{equation}
 T_{zz}=\frac{f^2}{2}(u')^2-f^2m^2(1-\cos u).
 \label{eq:appTzzu}
\end{equation}
Thus the stress obtained directly from the nonlinear higher-form action agrees with the corresponding Sine--Gordon expression.

\section{Long-wavelength expansion of the CSL phonon}
\label{app:phononlong}

The phonon dispersion of the CSL was derived in Ref.~\cite{Brauner:2016pko}.  Here we give an alternative derivation of its long-wavelength form, using the displacement field introduced in Sec.~6.2.  We start from
\begin{equation}
 -\frac{d}{dz}\left(\rho(z)\frac{d\xi}{dz}\right)
 =\varepsilon\,\rho(z)\xi(z),
 \label{eq:appSL}
\end{equation}
where $\rho(z)$ is periodic with period $\ell$.  At $p_z=0$, the lowest mode has $\varepsilon=0$ and $\xi=\text{const.}$, corresponding to a uniform translation of the CSL.

For nonzero Bloch momentum we impose
\begin{equation}
 \xi(z+\ell)=e^{ip_z\ell}\xi(z).
 \label{eq:appBloch}
\end{equation}
When $p_z\ell\ll1$, the lowest mode differs only slowly from the translational zero mode.  To leading order we write
\begin{equation}
 \xi(z)=e^{ip_z\theta(z)}+O(p_z^2),
 \qquad
 \theta(z+\ell)-\theta(z)=\ell.
 \label{eq:apptheta}
\end{equation}
Multiplying Eq.~\eqref{eq:appSL} by $\xi^*(z)$ and integrating over one period gives
\begin{equation}
 -\int_0^\ell dz\,\xi^*(z)
 \frac{d}{dz}\left(\rho(z)\xi'(z)\right)
 =
 \varepsilon\int_0^\ell dz\,\rho(z)|\xi(z)|^2.
\end{equation}
Integrating the left-hand side by parts, the boundary term vanishes because of the Bloch condition \eqref{eq:appBloch} and the periodicity $\rho(z+\ell)=\rho(z)$.  We therefore obtain
\begin{equation}
 \int_0^\ell dz\,\rho(z)|\xi'(z)|^2
 =
 \varepsilon\int_0^\ell dz\,\rho(z)|\xi(z)|^2.
\end{equation}
Substituting Eq.~\eqref{eq:apptheta} into this relation gives
\begin{equation}
 \varepsilon
 =p_z^2
 \frac{\displaystyle\int_0^\ell dz\,\rho(z)\left[\theta'(z)\right]^2}
 {\displaystyle\int_0^\ell dz\,\rho(z)}
 +O(p_z^4).
 \label{eq:appeps}
\end{equation}

For fixed Bloch momentum $p_z$, the lowest eigenvalue is obtained by minimizing the numerator with respect to $\theta(z)$.  The Bloch boundary condition requires
\begin{equation}
 \theta(\ell)-\theta(0)
 =\int_0^\ell dz\,\theta'(z)=\ell.
 \label{eq:appconstraint}
\end{equation}
Varying $\theta$ gives
\begin{equation}
 \frac{d}{dz}\left[\rho(z)\theta'(z)\right]=0,
\end{equation}
and therefore
\begin{equation}
 \theta'(z)=\frac{C}{\rho(z)},
 \qquad
 C=\frac{\ell}{\displaystyle\int_0^\ell dz\,\rho^{-1}(z)}.
 \label{eq:appC}
\end{equation}
It follows that
\begin{equation}
 \int_0^\ell dz\,\rho(z)\left[\theta'(z)\right]^2
 =\frac{\ell^2}{\displaystyle\int_0^\ell dz\,\rho^{-1}(z)}.
\end{equation}
Inserting this result into Eq.~\eqref{eq:appeps}, we obtain
\begin{equation}
 \varepsilon=c_z^2p_z^2+O(p_z^4),
 \qquad
 c_z^2=
 \frac{\ell^2}
 {\left(\displaystyle\int_0^\ell dz\,\rho(z)\right)
  \left(\displaystyle\int_0^\ell dz\,\rho^{-1}(z)\right)},
\end{equation}
which is Eq.~\eqref{eq:czgeneral}.


\end{document}